\documentstyle[11pt]{article}
\newcommand{\sect}[1] {\section{#1}\setcounter{equation}{0}}

\renewcommand{\theequation} {\thesection.\arabic{equation}}

\newcommand{\be}{\begin{equation}}
\newcommand{\BE}{\begin{equation}}
\newcommand{\BL}[1] {\begin{equation}\label{#1}}
\newcommand{\EE}{\end{equation}}
\newcommand{\BA}[1] {\begin{eqnarray}\label{#1}}
\newcommand{\EEA}{\end{eqnarray}}
\newcommand{\BAN} {\begin{eqnarray}\nonumber}
\newcommand{\bay}{\begin{array}}
\newcommand{\eay}{\end{array}}

\newcommand{\II}{I} \newcommand{\NN}{J}
\newcommand{\ZZ}{{}^{(1)}Y} \newcommand{\YY}{Y}
\newcommand{\FF}{{}^{(1)}I} \newcommand{\ZZZ}{Z}

\newcommand{\al}   {\alpha}

\newcommand{\ff}   {\varphi}
\newcommand{\dl}   {\delta}
\newcommand{\dz}   {\partial}

\newcommand{\kk}   {\kappa}

\newcommand{\lm}   {\lambda}

\newcommand{\ux}   {\underline x}
\newcommand{\uz}   {\underline z}
\newcommand{\IZ} {Z\!\!\!Z}

\begin{document}
\pagestyle{empty}
\title{\mbox{ }\\[19.2 mm]
 $A_1^{(1)}\,$ ADMISSIBLE REPRESENTATIONS --
 FUSION TRANSFORMATIONS AND LOCAL CORRELATORS}

\author{ P. Furlan\\ Dipartimento  di Fisica Teorica
 dell'Universit\`{a} di Trieste, Italy,\\
and Istituto Nazionale di Fisica Nucleare (INFN), Sezione
di Trieste, Italy,\\
 [3.8mm] A.Ch. Ganchev
\\Institute for Nuclear Research and
Nuclear Energy, 72 Tsarigradsko Chausee, 1784 Sofia, Bulgaria,
\\[3.8mm] and \\
\\[2.8mm] V.B. Petkova\\
Institute for Nuclear Research and
Nuclear Energy, 72 Tsarigradsko Chausee, 1784 Sofia,
Bulgaria{${}^*$},\\
and Istituto Nazionale di Fisica Nucleare, Sezione di Trieste,
Trieste, Italy}
  \date{}
\maketitle
\begin{abstract}

We reconsider the earlier found solutions of the
Knizhnik-Zamolodchikov (KZ) equations describing correlators
based on the admissible representations of  $A_1^{(1)}$.
Exploiting  a symmetry of the KZ equations we show that the
original infinite sums representing the 4-point chiral
correlators can be effectively summed up.  Using these simplified
expressions with proper choices of the contours we determine the
duality (braid and fusion) transformations and show that they are
consistent with the fusion rules of Awata and Yamada.  The
requirement of locality leads to a 1-parameter family of
monodromy (braid) invariants.  These analogs of the  ``diagonal''
2-dimensional local 4-point functions in the minimal Virasoro
theory contain in general non-diagonal terms.  They correspond to
pairs of fields of identical monodromy, having one and the same
counterpart in the limit to the Virasoro minimal correlators.

\end{abstract}

\footnotetext
{${}^*$ Permanent address. \\ \quad \\ {}\qquad
}

       \vspace{-20 cm} \hspace{11 cm}

       \hspace{8 cm}

\textheight =22.08cm

\thispagestyle{empty}
\newpage
\setcounter{page}{1}
\pagestyle{plain}

\newpage
\sect{Introduction}

The WZNW models based on the set of integrable representations
of an affine Lie algebra $\hat g$ at integer level $k$ are the best
studied conformal models. This is so because on one hand
more complicated models can be obtained from them by the coset
construction and on the other hand they are simple enough
and the full description is probably within reach.
\footnote{The operator product coefficients
(except for the case of $g=su(2)$) are still unknown
and the classification of the
modular invariants is not yet completed.}
The simplicity of these models is most apparent in the equations
governing their correlators --- the Knizhnik-Zamolodchikov system
of equations \cite{KZ}, which has been solved in the general
case of integrable
representations \cite{SV} (see also the earlier works \cite{ZF},
\cite{ChF}, \cite{DJMM} for the case $\widehat{sl(2)}_k)$.

Another class of examples is provided by the admissible
representations
for rational levels (the integrable ones being a particular case
of the admissible when the level is a positive integer) introduced
and studied in depth by \cite{KW}. The study of WZNW conformal
models based on admissible representations (see \cite{Gaw} for a
lagrangean approach in the case of rational level) is motivated
also by the fact that from
these models one can obtain others by quantum hamiltonian reduction
\cite{BO}.

In a series of papers \cite{FGPP}, \cite{FGPP2} we have studied the
reduction  of the admissible $\hat{sl}(2)$ models to the Virasoro
minimal models on the level of correlators. The first difficulty in
describing the correlators of admissible fields comes from the fact
that the underlying representations of $sl(2)$ characterised by
rational isospins are infinite dimensional, in general  neither
lowest nor highest weight representations. In the free field
approach
this reflects in the necessity to consider screening currents
involving rational powers of the ghost fields and a first attempt
in this direction was made in \cite{Do}. This difficulty we have
overcome by passing to a functional realization of $sl(2)$ in terms
of an ``isospin coordinate''  $x$ \cite{ZF} and expanding the
correlators in an infinite power series in the differences
$(x_a-z_a)$
with coefficients that are $z$ dependent multiple contour integrals
($z_a$ being the ``space'' coordinates). In this expansion the
quantum
reduction becomes immediate since the zero order coefficient
coincides
(for generic values of the isospins) with the minimal model
correlators
of Dotsenko and Fateev (DF) \cite{DF}; thus the reduced theory is
obtained by setting $x_a=z_a$. The numerical coefficients in the
expansion are fixed requiring that the correlators satisfy the KZ
equations and the Ward identities accounting for the
$sl(2)\oplus sl(2)$
invariance. (Alternatively  the result was reproduced in the free
field
approach exploiting a particular way of giving
meaning to the second screening charge.)

Other approaches to the problem have been proposed \cite{FMI},
\cite{PRY}, \cite{And}, leading to different integral
representations
for the correlators.  The relation between them is still not very
clear.
There is another important issue -- namely the braiding properties
of the  correlators based on the admissible representations, which
is not yet systematically investigated. A drawback of our old
representation was that the infinite in general power series in
$x-z$ hides a nontrivial analytic behaviour around $x$. A naive
term by  term analysis has led us to an incorrect statement about
the fusion rules of admissible fields. After these fusion rules
have been found by different means by Awata and Yamada \cite{AY}
(see also \cite{FM2} and the recent paper \cite{FM}) the question
of their consistency with the analytic properties of the
correlators still remains open.

We address this problem in the present paper continuing the program
started in our earlier work. Recalling in Section 2 some basic
definitions and the main results about the solutions of the KZ
equations found in \cite{FGPP}, \cite{FGPP2} (see also \cite{GP}),
we then show in Section 3 that the infinite sums in the powers of
$(x-z)$ representing the 4-point correlators can be effectively
summed up. The resulting simple expressions are precisely the
chiral counterparts of the corresponding 2-dimensional physical
4-correlators proposed recently in \cite{And} as a generalisation of
the construction of \cite{ZF}. Our next step in Section 4 is to
select a basis in the space of 4-point blocks, characterised by a set
of contours (cycles), extending  the one in \cite{DF}. The blocks
reproduce the contribution of all fields described by the fusion
rules of Awata and Yamada. Furthermore in Section 5 we show that
this set is
closed under braid (fusion) transformations, when exchanging pairs of
coordinates $(x_a\,,\,z_a)$, and explicitly compute the matrix
elements
of these transformations. We point out the existence of another basis
of integrals such that a subset of it is closed under the standard
fusion transformations  of the minimal theory, given (up to signs) by
a product of two $U_q(sl(2))$ $6j$-symbols. These results are used to
construct  local 4-point functions. We find basically two types of
monodromy (braid) invariants, which can be looked as analogs of the
diagonal DF  physical correlators recovered in the limit $x \to z$.
A particular linear combination of these two basic invariants is
identical to the volume integral representation proposed by Andreev
in \cite{And} and allows to complete the computation of the
corresponding OPE structure constants. The general monodromy
invariants however, though containing as labels all the ``exponents''
in the theory, are not ``diagonal'' and  involve mixed terms.
The latter correspond to fields with different values of the isospin
label $J$  and the Sugawara conformal weight $\Delta_J$, but
identical
reduced weight $h_J=\Delta_J-J$, which have effectively the same
monodromy. This fact is alternatively reflected in the existence of
yet another physical $4$-point functions with mixed left and right
chiral content, corresponding to chiral correlators of
the Virasoro minimal and the rational level WZNW models respectively.
The technical details of the computations are collected in an
Appendix.

%
%
\sect{Solutions of the KZ equations with rational isospins}

Recall the admissible representations of $A_1^{(1)}$  \cite{KW}
labelled by a rational level
\BL{rat}
   k+2=p/p' \equiv \kk  \,,
\EE
($p,p'$ -- coprime (positive) integers), and isospins
\BL{kJ}
      2J_{r,r'}=r-r' \kk= 2j-  2j'\,\kk \,,
\EE
where $\,r, r'$ are nonnegative integers, restricted furthermore
to
\BL{jj'}
  0 \le r \le p-2, \quad 0 \le r' \le p'-1\, .
\EE
Let $r'\not = p'-1$. Upon quantum hamiltonian reduction the
representations labelled by $J=J_{r,r'}\,$ and
\BL{afW}
  J^{(1)} = \kk-J-1\,, \quad
  J_{r,r'}^{(1)} = J_{p-r-2,p'-r'-2}\,,
\EE
both reproduce the Virasoro irreducible representations
characterised
by a central charge $c_k=13-6 \kk -6/\kk\,$ and conformal weight
\BL{h(J)-J}
  h_J = \Delta_J -J =h_{ J^{(1)} }\ ,
\EE
where $\Delta_J$ is the Sugawara scale dimension
\BL{sugaw}
   \Delta_J={J(J+1) \over k+2 }=   \Delta_{-J-1}\ .
\EE
The quantum fields $\Psi_{J}(z,x)$ depend on two complex
variables and  transform with respect to the semidirect product
of $A_1^{(1)}$ and the Virasoro algebra according to
\BL{trV}
  [L_n, \Psi_{J}(x,z)] = ( z^{n+1}\partial_z
                    + (n+1) \Delta_J\,z^n)\, \Psi_{J}(x,z)\,,
\EE
\BL{tr}
  [X_n^{\al}, \Psi_{J}(x,z)] = z^n\, S^{\al}\, \Psi_{J}(x,z)\,,
\EE
\BL{hl}
   S^0=-2x \dz_x + 2J,\quad  S^-=-\partial_x,
   \quad  S^+=x^2\dz_x-2Jx \,.
\EE
For real $x$ (\ref{hl}) are the generators of an (infinite
dimensional) induced representation of $SL(2,R)$.
Whenever  $2J$ is a nonnegative integer the representation
space has a finite dimensional  invariant subspace
described by polynomials of $x$ of maximal degree $2J$ .
Alternatively we can look at them as the generators resulting
after analytic continuation to noninteger $J$ of the induced
representations of type $(J,0)$ of the algebra $sl(2,C)$.

The state $|J\rangle= \Psi_J(0,0)|0\rangle$ represents a highest
weight state (h.w.s.) of a Verma module with respect to the
generators  $X_n\,, L_n$. Similarly
$\Psi_J(x,z)|0\rangle=e^{z\, L_{-1}-x\, S^-}\,|J\rangle\,$
can be viewed as a Verma module h.w.s.
with respect to the generators
\BL{raX}
  {\cal X}_n^{\al}(x,z)\, \Psi_J(x,z)|0\rangle
  := e^{z\, L_{-1}-x\, S^-}\, X_n^{\al}\,|J\rangle\,,
\EE
\BL{raL}
  {\cal L}_n(z)\, \Psi_J(x,z)|0\rangle
  := e^{z\, L_{-1}-x\, S^-}\, L_n\,|J\rangle\,.
\EE

The chiral correlators of the fields $\Psi_{J}(x,z)$ are
invariant
with respect to the $sl(2,C) \times sl(2,C)$ subalgebra of
(\ref{trV}), (\ref{tr}).
They furthermore satisfy the KZ system of equations
\BL{KZ}
   \left( \frac{\dz}{\dz z_a} - {1 \over \kk}
   \sum_{{}^{b=1}_{b \ne a}}^n
   \frac{\Omega_{ab}}{(z_a-z_b)}\right)
  {W}^{(n)} (x_1,z_1,J_1;\dots;x_n,z_n,J_n) = 0 \, ,
  \quad a=1,2,....n\,,
\EE
\BL{Omega}
  \Omega_{ab}=
  S^{\al}_{a}\, S_{b,\al}
= - x_{ab}^{\,2}\, {\dz\over\dz x_a}\,
  {\dz\over\dz x_b}
  + 2 x_{ab}\left( J_a {\dz\over\dz x_b}
  - J_b  {\dz\over\dz x_a} \right) + 2J_a J_b \, ,
\EE
alternatively rewritten as
\BL{KZr}
  \Big({\cal L}_{-1,a} - {1\over \kk}\,( {\cal X}^+_{-1,a}\,
  {\cal X}^-_{0,a} +
  {1\over 2}\, {\cal X}_{0,a}^0 \, {\cal X}^0_{-1,a})\Big)\,
  {W}^{(n)} (x_1,z_1,J_1;\dots;x_n,z_n,J_n) = 0 \,.
\EE
Here for $n \ge 0$
\BL{cX}
  {\cal X}_{-n,a}^-=- \sum_{b (\not = a)}\, {S_b^-\over
  z_{ba}^{n}}\,, \ \
  {\cal X}_{-n,a}^0= -\sum_{b (\not = a)}\,
  {-S_b^0+2x_a\,S_b^-\over z_{ba}^{n}}\,, \ \
  {\cal X}_{-n,a}^+=- \sum_{b (\not = a)}\,
  {S_b^+ + x_a\,S_b^0-x_a^2\,S_b^-\over  z_{ba}^{n}}\,.
\EE
\BL{cL}
  {\cal L}_{-n,a}= \sum_{b (\not = a)}\, {1\over z_{ba}^{n-1}}\,
  \Big({(n-1) \Delta_{J_b}\over z_{ba}} -{\dz\over \dz z_b}\Big)\,.
\EE
The Ward identities imply that
\BL{WX}
 {\cal X}_{0,a}^+\, W=0\,,\qquad  {\cal X}_{0,a}^0\,
 W=2J_a\,W\,, \qquad  {\cal L}_{0,a}\,W=\Delta_{J_a}\, W\,, \,
\EE
(so that in (\ref{KZr}) ${\cal X}_{0,a}^0\,$   can be replaced by
$2J_a$) and furthermore
${\cal X}_{0,a}^-\,W = S_a^-\,W =-{\dz \over \dz x_a}\,W\,,$
${\cal L}_{-1,a}\,W =L_{-1,a}\,W= {\dz \over \dz z_a}\,W$. (Here
$W=W^{(n)}\,.$)
The operators (\ref{cX}) (\ref{cL}) (see \cite{BPZ}, \cite{ZF})
are derived from  (\ref{raX}), (\ref{raL}) and their
generalisations for arbitrary number of operators\footnote{
E.g., the generalisation of   (\ref{raX}) reads
$$
 {\cal X}_n^{\al}(x,z)\, \Psi_J(x,z) \, \Psi_{J_r}(x_r,z_r)\,
 \cdots
 \Psi_{J_1}(x_1,z_1)\, |0\rangle := e^{z\, L_{-1}-x\, S^-}\,
 \{[  X_n^{\al}, \Psi_J(0,0)] \,
 \Psi_{J_r}(x_r-x,z_r-z)\, \cdots
$$ $$ \cdots
 \Psi_{J_1}(x_1-x,z_1-z) + \Psi_J(0,0)\, \Psi_{J_r}(x_r-x,z_r-z)
 \, \cdots
 \Psi_{J_1}(x_1-x,z_1-z)\,  X_n^{\al} \}\,|0\rangle\,,
$$ etc.},
moving $X_{-n}$ (or $L_{-n}$) to the left and using the
commutators
(\ref{tr}), (\ref{trV}). By the definition in  (\ref{raX}),
(\ref{raL}) the positive mode generators  vanish.

The correlators furthermore should satisfy two series of algebraic
equations resulting from the presence of singular vectors in the
Kac--Moody Verma modules labelled
by $J=J_{rr'}$ $\equiv J_{r-p+2,r'-p'+2}\,,$
\BL{MFF}
  {\cal P}_i({\cal X}_{-1,a}^+, {\cal X}_{0,a}^-)\, W=0\,,
\EE
where  ${\cal P}_i(X_{-1}^+, X_0^-)\,, i=1,2\,,$  are some
monomials
in powers of the Kac--Moody generators $X_{-1}^+, X_0^-\,,$
\cite{MFF}, realised here in terms of the generators (\ref{cX}).
If $J_a$ is of the type $J_{m,1}\,,$ the first of  these operators
is simply $({\cal X}_{0,a}^-)^m= (-\dz_{x_a})^m\,.$

Let us now describe the explicit form of the solutions found
in \cite{FGPP},
\cite{FGPP2}. Up to a standard prefactor the full $n$-point
correlators are recovered from the correlators ``at infinity''
$$
  W^{(n,\infty)}(\{x_a,z_a,J_a\}):=\lim_{x_n,z_n \to \infty}\,
  x_n^{-2 J_n}
  \,z_n^{2\Delta_{J_n}}\,\langle 0|\,\Psi_{J_n}(x_n,z_n) \cdots
  \Psi_{J_1}(x_1,z_1)\,|0\rangle
$$
$$
  = \langle J_n |\, \Psi_{J_{n-1}}(x_{n-1},z_{n-1})
  \cdots \Psi_{J_1}(x_1,z_1)|0\rangle
$$
if furthermore $x_a\,, z_a$ are replaced by
$$
\ux_a = \frac {x_{1a} \, x_{n-1,n}}{x_{1,n-1}\,
x_{an}}\,, \qquad
  \uz_a = \frac {z_{1a} \, z_{n-1,n}}{z_{1,n-1}\, z_{an}}\,,
\qquad    a=1,2,\dots,n-1\ .
$$
In the counterpart of (\ref{KZ}) applied to $W^{(n,\infty)}$
the indices $a,b$ run over $1,2,...,n-1$; the same applies to the
generators (\ref{cL}), (\ref{cX}), now defined for $n>0\,$.
The zero mode subalgebra acts as in (\ref{WX}) and the surviving
Ward identities for $X_0^0$ and $L_0$ get accordingly modified by
the values $J_n\,, \Delta_{J_n}\,$ at infinity
(see e.g., \cite{GP}).
The operators ${\cal P}_i$   can be also realised in terms of the
generators $\triangle(X_{-1}^-)\, $ and $\triangle(X_{0}^+)$,
(replacing $X_{-1}^+\,$ and $X_0^-\,$ respectively),
where $\triangle(X_{n}^{\alpha})=\sum_{a=1}^{n-1}\,
X_{n,a}^{\al}\,, $
resulting from the action on the dual vacuum state $\,\langle
J_n|$.

Now assume that the real numbers $J_a=j_a-j'_a\,\kk\,, \,
a=1,2,...,n\,$ (with $2j_a, 2j'_a\,$ not necessarily
satisfying (\ref{jj'})) are such that both $s$
and $s'\,, $ defined according to
$$
  s^{(')}=\sum_{a=1}^{n-1}\,j^{(')}_a-j^{(')}_{n}\,,
$$
are nonnegative integers. Denoting
\BL{cha}
  S=\sum_{a=1}^{n-1}\, J_a-J_{n}\,=s-s'\,\kk\,,
\EE
the correlator $W^{(n,\infty)}$  reads
\BL{x-zn}
  W^{(n,\infty)}(\{x_a,z_a,J_a\})
  =\prod_{1\le b<a\le n-1}(z_{a}-z_b)^{2J_aJ_b/\kk}\,
    \sum_{t=0} \sum_{\tau:\,\,|\tau|=t,}  \prod_{a=1}^{n-1}
   (x_a- z_a)^{\tau_a}
   B_{\tau}\ I^{(s,s')}_{\Gamma, \tau}(\{z_a,J_a\})\ ,
\EE
where $\tau$ is a $n-1$ vector with nonnegative integer
components $t_a$, $B_{\tau}$ are numerical coefficients
\BL{B_tau2}
  B_{\tau}=\beta_{J}\,\frac{(-1)^t\,\Gamma(c_t+1)}{\Gamma
  (2J_n+2-\kk)\Gamma(S-t+1)}
  \ \prod_{a=1}^{n-1} {2J_a\choose \tau_a},\quad t=|\tau|=
  \sum_{a=1}^{n-1}\, \tau_a\, ,
\EE
where ${a \choose k}={\Gamma(a+1)\over \Gamma(a-k+1)\,k!}\,$
and $\beta_{J}$ is an  overall constant,
\BL{tc}
  c_t=c_{t}(J)=\sum_{a=1}^{n}J_a - t +1 - \kk=S-t+2J_n+1-\kk\, ,
\EE
while $ I^{(s,s')}_{\Gamma, \tau}\,$ are contour integrals
\BL{nint}
   I_{\Gamma,\tau}^{(s,s')}(\{z_a,J_a\})=
   \int_{\Gamma} du_1\dots du_s
   \,  dv_1\dots
   dv_{s'} \, \,  \ff_{\tau}^{(s,s')}(u_i,v_{i'};z_a) \,
   \Phi_{{J}}^{(s,s';\kk)}(u_i,v_{i'};z_a)\, .
\EE
The factor $\Phi_{{J}}^{(s,s';\kk)}$ is the standard integrand
of the DF conformal blocks,
\BL{PhiPhi}
   \Phi^{(s,s';\kk)}_{{J}}(u_i,v_{i'};z_a) =
   \Phi^{(s;1/\kk)}_{{J}}(u_i;z_a)\,
   \Phi^{(s';\kk)}_{- {J}/\kk}(v_{i'};z_a)
   \prod_{i=1}^s\prod_{i'=1}^{s'} {1\over (u_i-v_{i'})^2}\, ,
\EE
and
\BL{Phi}
   \Phi^{(s;1/\kk)}_{{J}}(u_i;z_a) = \prod_{j<i}^s (u_i-u_j)^{2/\kk}
   \prod_{i=1}^s\prod_{a=1}^{n-1} (u_i-z_a)^{-2J_a/\kk} \, .
\EE

To describe the factor $\varphi^{(s,s')}_{\tau}$ introduce
the $(n-1)$-vector $\mu$ with non-negative integer components
$\mu_a\,,$ and set for any integer $m$ and  arbitrary $L$
\BL{Lphi}
   \varphi_{\mu}^{[L,m]}(\{u_i,z_a\}) =
  {\Gamma(mL- |\mu|+1) \over \Gamma(mL+1)}
  \sum_{\{N_{ia}\}}\,
  \prod_{i=1}^m\,{\Gamma(L+1)\over
  \Gamma( L-\sum_{a=1}^{n-1}\,N_{ia}+1)}\,
  {\prod_{a=1}^{n-1}\, [\mu_a! \, \dl(\sum_{i=1}^{s}
  N_{ia} -\mu_a) ]   \over \prod_{i=1}^m\,
  \prod_{a=1}^{n-1}\,  [N_{ia}!(u_i-z_a)^{ N_{ia}}]} \ .
\EE
The sums in (\ref{Lphi})  are finite, running
over $\ N_{ia}=0,1,...,\mu_a\ ,$ subject to the
constraints $\sum_i N_{ia} = \mu_a \,.$
For $L=1\,$  $N_{ik}\,$ get furthermore restricted
to  $N_{ik}=0,1$. In general (\ref{Lphi})  is used to define
the factor $\varphi^{(s,s')}_{\tau}(\{u_i,v_{i'},z_a\})$
in (\ref{nint}) according to \cite{FGPP}, \cite{FGPP2}
\BL{nphi}
   \varphi^{(s,s')}_{\tau}(\{u_i,v_{i'},z_a\}):={1
   \over {S\choose s} }\, \sum_{\lm }\,
   \prod_{a=1}^{n-1}\, {\tau_a \choose \lm_a} {S-|\tau| \choose
   s-|\lm|}
   \varphi_{\tau-\lm}^{[-\kk,s']}(\{v_{i'},z_a\})
   \varphi_{\lm}^{[1,s]}(\{u_i,z_a\}) \ ,
\EE
where the sum  runs
over $\{\lm_a=0,1,..., {\rm min}(s, \tau_a)\},
\quad |\lm| \leq {\rm min}(s,|\tau|)$. Apparently for
any $\tau$ (\ref{nphi}) is a meromorphic function
of $\{u_i, v_{i'}, z_a \}\,.$

In the limit
$x_a\to z_a\,,$ which was related to the QHR \footnote{
This limit was generalised and applied  to the 2-point
functions of the $sl(n)$ WZNW model in \cite{FGPP3}. },
(\ref{x-zn}) (taken with the standard DF  \cite{DF}
contours $\Gamma\,$)  recover for generic spins the general
DF  correlators  of the minimal Virasoro theory, represented
by the $0^{{\rm th}}$ term in the (infinite) expansion
(\ref{x-zn}).
For $s'=0$ the sums in (\ref{x-zn}) become finite.

Although a direct check of the general algebraic equations
(\ref{MFF}) is still missing for the above correlators, it was
argued
in \cite{GP} that their validity is related to the fact that the DF
counterparts satisfy the BPZ equations, resulting from the singular
vectors in the corresponding Virasoro Verma modules. The latter
equations are expressed in terms of Virasoro generators realised by
operators similar to (\ref{cL}), with the Sugawara dimension
$\Delta_J$ replaced by $h_J\,.$  These arguments were further
supported by the proof, done within the algebraic BRST framework,
that the MFF singular
vectors reproduce the corresponding Virasoro ones modulo trivial
terms in the $Q_{\rm BRST}$ cohomology \cite{FGP}.

Applied to the $3$-point functions the equations (\ref{MFF}) yield
the fusion rules, computed explicitly in this way in  \cite {AY}.

%
%
\sect{Simplified expression for the 4-point correlators due to a
symmetry of the KZ equations.}

We now turn to the case of 4-point correlators. In the
system $(0,z=\uz_2,1)\, $ the summation in (\ref{x-zn}) reduces to
a single sum, $\tau=(0,t,0)\,,$ and the  KZ equation applied to the
series (\ref{x-zn}) turns to an infinite set of equations for the
integral coefficients $I^{(s,s')}_{\Gamma,t}(J;z):=
I^{(s,s')}_{\Gamma,(0,t,0)}(\{\uz_a,J_a\})\,$
\BL{ZF}
   ( \kk \dz_z+a_{t}(J,z)\, )\ I^{(s,s')}_{\Gamma,t}(J;z)
  +{t \ c_{t-1}(J)\over z(z-1)}\, I^{(s,s')}_{\Gamma,t-1}(J;z)
   = (2J_2 - t)\, (S-t)\ I^{(s,s')}_{\Gamma,t+1}(J;z)  \ ,
\EE
where
\BL{a}
    a_{t}(J,z)= {t(2J_1+2J_2-t+1) \over z}+{t(2J_3+2J_2-t+1)
        \over z-1},
    \quad {\rm and}\quad t=0,1,2,... \,.
\EE
Introduce the new set of isospins
\footnote{There is  a misprint on page 683 of \cite{FGPP2}
in the analogous formula for $\tilde{j}_4$ ($j_1$ and $j_4$
should be reversed), as well as in eq. (3.14),
in which $c_t$ should be replaced by $c_{t-1}$.}
\BA{tJ}
 \tilde {J}_1 & =& {1 \over 2}(J_1+J_2-J_3-J_4-1+\kk) \,,
\\
 \tilde {J}_2 & =& {1 \over 2}(J_1+J_2+J_3+J_4+1-\kk) \,,
\\
 \tilde {J}_3 & =& {1 \over 2}(-J_1+J_2+J_3-J_4-1+\kk) \,,
\\ \label{tt}
 \tilde {J}_4 & =& {1 \over 2}(-J_1+J_2-J_3+J_4-1+\kk) \,.
\EEA
Note that $J_i\to\tilde J_i$ is an involution. This change  of
variables (\ref{tJ}) keeps invariant $s$ and $s'$ as well as
the numerical coefficient $a_t(J,z)\,$ in (\ref{a}) since
$\tilde {J}_1+\tilde {J}_2=J_1+J_2\,,$ $ \tilde
{J}_3+\tilde {J}_2=J_3+J_2\,.$
Replacing  everywhere in (\ref{ZF}) the isospins $J_a$
by $\tilde{J}_a$ as defined in (\ref{tJ}) we can identify the
set of integrals $I_{\Gamma,t}^{(s,s')}(\tilde {J};z)\,$ serving
as solutions of the new equation with
\BL{I'}
  I_{\Gamma,t}^{(s,s')}(\tilde {J};z)={\rm const}\,
  (-1)^t\,
  \frac{B_t}{{S\choose t}}\, I_{\Gamma,t}^{(s,s')}(J;z)\,,\qquad
  t=0,1,2,... \,,
  \qquad B_t:=B_{(0,t,0)} \ ,
\EE
since the equation (\ref{ZF}) remains invariant under this change.
To see this multiply (\ref{ZF})  by the $t$-dependent constant
in (\ref{I'}) and use that
$$
  \frac{B_t}{{S\choose t}}\, c_{t-1}(J)=
  \frac{B_t}{{S\choose t}}\, (2\tilde{J}_2-t+1)
  = \frac{B_{t-1}}{{S\choose t-1}}\,\,  c_{t-1}(\tilde J) \,,
$$ $$
  \frac{B_t}{{S\choose t}}\, (2J_2-t)
  = \frac{B_{t+1}}{{S\choose t+1}}\,\, ( 2\tilde{J}_2-t) \,.
$$

 For $t=0$  (\ref{I'}) reduces to a relation for the DF
integrals of the minimal Virasoro theory. This in particular
allows to compute the overall, $t$-independent, constant
in (\ref{I'}) --  taking $t=0$ and using the
results of \cite{DF} for the normalisation
constants of the integrals $I_{\Gamma,0}^{(s,s')}(J;z)$ and
$I_{\Gamma,0}^{(s,s')}(\tilde J;z)$. Note that due to the
invariance of the combinations $J_1+J_2\,$ (or $J_3-J_4$), $
J_2+J_3\,$ (or $J_1-J_4$)   under the transformation  (\ref{I'}),
and due to the relation $\tilde{J}_2-\tilde{J}_4 +
\kappa-1=J_1+J_3\,$
(or  $\tilde{J}_1-\tilde{J}_3=J_2-J_4 +\kappa-1\,$),
these two types of integrals admit an identical asymptotic
behaviour for a fixed contour in the sets corresponding to
any of the $s- \,,$ $t- \,,$ and $u$-channels.

The observation about the invariance of the KZ-ZF system of
equations (\ref{ZF}) under the transformation (\ref{I'})  was made
in  \cite{FGPP2} in the particular case $s'=0$ in order to compare
with the solutions of the system as presented in \cite{ZF}, where
the combinations of isospins appear. Applied however in the more
general case $s'\not =0$ it allows to effectively sum the infinite
power in the 4-point function (\ref{x-zn}). Indeed for
$\tau=(0,t,0)\,$
the factor (\ref{nphi}) simplifies to
$$
  \varphi^{(s,s')}_{(0,t,0)}(\{u_i,v_{i'},z\})
  = \frac{1}{{S\choose t}}
  \, \sum_{l=0}^{{\rm min}(s,t)}\,
  \sum_{\{N_{i'}'\}} \,\dl(\sum_{i'}^{s'}\, N_{i'}' -t+l)\,
  \prod_{i'=1}^{s'}\, {-\kk \choose N_{i'}'}\,
  {1  \over (v_{i'}-z)^{ N_{i'}'} } \,\,
$$
$$
  \sum_{\{N_{i } \}} \,\dl(\sum_i^s \, N_i -l)\,
  \prod_{i =1}^s \, {1 \choose N_{i } }\,
  {1 \over (u_i -z)^{N_i} }\,,
$$
and it is straightforward  to show that
$$
  \varphi^{(s,s';\kk)}(\{u_i,v_{i'};x,z\}):=   \sum_{t=0}^{\infty}
  (-1)^t\,  {S \choose t}\,   ( x- z)^{t}\,
  \varphi^{(s,s')}_{(0,t,0)}(\{u_i,v_{i'},z\})
$$
\BL{sphi}
   = \prod_{i=1}^s\, \Big(  \frac{u_i-x}{u_i-z} \Big)
    \,\prod_{i'=1}^{s'}\, \Big(  \frac{v_{i'}-x}{v_{i'}-z}
  \Big)^{-\kk}\,.
\EE
Hence  taking into account (\ref{I'}) we can rewrite the 4-point
correlator (\ref{x-zn}) as
\BL{scor}
  W^{(4,\infty)}(x,z,J)= {\rm const}\,
  z^{2J_1 J_2/\kk}\, (1-z)^{2J_3 J_2/\kk}\,
  I^{(s,s')}_{\Gamma}(\tilde J;x,z)\,,
\EE
\BL{sint}
 I^{(s,s')}_{\Gamma}(\tilde J;x,z): =
 \int_{\Gamma} du_1\dots du_s
 \,dv_1\dots   dv_{s'} \,\,
 \varphi^{(s,s';\kk)}(\{u_i,v_{i'};x,z\})\,
 \,\Phi_{\tilde J}^{(s,s';\kk)}(u_i,v_{i'};z)\, .
\EE
In the limit $x\to z$ (\ref{scor}) provides an alternative
expression for the  DF conformal blocks in which in the
integrand $J_a$ are replaced by $\tilde{J}_a$.

The expression  (\ref{scor}), (\ref{sint}) extends the infinite
sum representing the correlator beyond its range of convergence.
(Expanding instead in powers of $(u_i-z_a)/(x_a-z_a)\,$
we recover the solutions with inverse powers of $(x_a-z_a)\,,$
mentioned in \cite{FGPP2}.) The explicit singularity for
noninteger $\kappa$ of the new factor (\ref{sphi}) suggests that
the set of contours (cycles) considered in \cite{DF} has to be
enlarged.
Our objective in the next section will be to study the braiding
properties of the integrals $I^{(s,s')}_{\Gamma}(\tilde J;x,z)\,$
for some choices of the contours $ \Gamma\,.$
\medskip

\noindent
{\bf Remark: }
The infinite sum in (\ref{x-zn}) can be taken also in a different
way for special values of the isospins. E.g., for $n=4\,,$ $s'=1$
and $2J_4+1=\kk$ we have $c_t=S-t$, the constant $B_t$ simplifies
to $B_t=(-1)^t\, {2J_2\choose t}\,,$
and the infinite sum  produces $({v-x\over v-z})^{2J_2}$, without
other changes in the integrand. In this form this example
coincides with the corresponding integral in \cite{FMI}. However,
unless we misunderstand the prescription there, this
identification holds under different restrictions on the isospins --
the condition (\ref{cha}) above, $J_1+J_2+J_3-J_4=-\kappa$
seems to be replaced in \cite{FMI} by $J_1+J_2+J_3+J_4+1=0$ and
the two coincide only for the special choice $2J_4+1= \kappa$.

\sect{Basic set of chiral blocks and fusion rules}
%
%

Recall the well known fusion rules of the WZNW
conformal field theory for  $A^{(1)}_1$  at integer level $p-2$.
If the highest weights of the integrable representations are
labelled  by $r(\equiv 2j)=0,1,\dots,p-2$ then the fusion
multiplicities $N^{r}_{r_1r_2}(p)$ are equal to $1$ if
$$
  r\in\{ |r_1-r_2|\,, |r_1-r_2|+2,\dots\,, p-2-|r_1+r_2-p+2|\}
$$
and  $N^{r}_{r_1r_2}(p)=0$ otherwise. The algebra of
matrices $N_r(p)\,,$ $(N_r)_{r_1}^{r_2}(p)=N_{r\, r_1}^{r_2}(p)$
is $\IZ_2$ graded, namely if $r\longmapsto \tau(r)=0,1\,$
if $r$ -- even,  odd, respectively, and if $N_{r_1r_2}^r=1$
then $\tau(r_1)+\tau(r_2)=\tau(r) \,{\rm mod}\, 2$,
and thus it possesses a subalgebra $\{N_r(p)\,,\tau(r)=0\}$.
Note that $N_{r\, p-r_1-2}^{p-r_2-2}(p)=N_{r\, r_1}^{r_2}(p)\,.$
We shall furthermore use the notation $J^{(t)}\,, t=0,1$ mod
$2\,,$ where $J^{(0)}\equiv J\,, $ and $J^{(1)}_{r r'}=
J_{p-r-2, p'-r'-2}\,,$
for $r'\not =p'-1\,,$ as defined  in (\ref{afW}) (i.e., the weight
obtained from $J$ by the shifted action of the affine
root Weyl reflection $w_0$).

The fusion rules for the admissible representations, first derived
by Awata-Yamada \cite{AY} can be written in the following form,
suggested by the consideration in the recent paper \cite{FM}
\BL{frs}
  J_{r_1\,,\,r'_1}\otimes J_{r_2\,,\,r'_2} =
  \sum_{r,r'}\, N^{r}_{r_1\,r_2}(p)
  N^{2r'}_{2r'_1\,2r'_2}(2p')\,
  J_{r\,,\,r'-\tau(r'_1+r'_2-r')}^{(\tau(r'_1+r'_2-r'))}\,,
\EE
where $N^{2r'}_{2r'_1\,2r'_2}(2p')$ are the fusion coefficients of
the fusion subalgebra $\{N_m(2p')\,, \,\tau(m)=0\}$  of a
$A^{(1)}_1$ theory at level $2(p'-1)\,.$  Thus in (\ref{frs}) $r'$
runs along all integers between $|r'_1-r'_2|$ and
$p'-1-|p'-1-r'_1-r'_2| \le p'-1\,.$
Both multiplicities $N^{r}_{r_1r_2}(p)$
and $N^{2r'}_{2r'_1\,2r'_2}(2p')\,, $ can be expressed
by a Verlinde type formula in terms of the modular matrices at
level $p-2$ and $2p'-2$ respectively.
In \cite{FM} Feigin and Malikov gave a ``supersymmetric''
interpretation of the fusion rule of Awata and Yamada,
showing that it coincides with the product of the  truncated
tensor products of $U_q(sl(2))$ and $U_{q'}(osp(1|2))$
for $q^p=1\,,(q')^{p'}=1\,,$  respectively. Note that
in (\ref{frs}) the sum in $r$ runs by two while the sum in $r'$ --
by one, the new terms counted by the ``odd'' isospins $J^{(t=1)}$.

Now we turn to the basis of 4-point conformal blocks.
Such a basis can be labelled by the isospin running in the
intermediate channel or equivalently is given
by an appropriate choice of the integration contours ${\cal C}$.

The $s$-channel  basis  is labelled by an isospin $J$ that is the
result of the fusion of $J_1$ and $J_2$. The $t$-channel
basis is labelled by an isospin $J$ that is the result of the fusion
of $J_3$ and $J_2$, and the $u$-channel --- by isospins $J$
in the fusion of  $J_4$ and $J_2$. The dimension of any of them is
$$
 \sum_{rr'} N^{(r\,r')}_{(r_1\,r'_1)(r_2\, r'_2)}
            N^{(r_4\,r'_4)}_{(r\,r')(r_3\,r'_3)} =
 \sum_{rr'} N^{(r\,r')}_{(r_3\,r'_3)(r_2\,r'_2)}
            N^{(r_4\,r'_4)}_{(r\,r')(r_1\,r'_1)} =
 \sum_{rr'} N^{(r\,r')}_{(r_1\,r'_1)(r_3\,r'_3)}
            N^{(r_4\,r'_4)}_{(r\,r')(r_2\,r'_2)} \,,
$$
where
$$
  N^{(r\,r')}_{(r_1\,r'_1)(r_2\, r'_2)}:
  = N^{r}_{r_1\,r_2}(p)\,  N^{2r'}_{2r'_1\,2r'_2}(2p')\,.
$$

We describe in more details the ``even'' part of the  $s$-basis.
Let  $m,m'$ be a pair of nonnegative
integers, $m=0,1,..., s\,, m'=0,1,...,s'\,.$
The ``even'' $s$-channel 4-point blocks (\ref{sint}) are given by
\BA{II}
  \II_{mm'}(a,b,c,d;x,z)
& = &
 \int_{{\cal C}_{m,m'}}
  \prod_{i=1}^m u_i^{-a/\kappa}
  (z-u_i)^{-1-c/\kappa}
  (x-u_i)(1-u_i)^{-b/\kappa} \nonumber\\
&  &\prod_{i=m+1}^s u_i^{-a/\kappa}
  (u_i-z)^{-1-c/\kappa}
  (u_i-x)(u_i-1)^{-b/\kappa} \\
&  &  \prod_{i'=1}^{m'} v_{i'}^{a}
  (z-v_{i'})^{c+\kappa}
  (x-v_{i'})^{-\kappa}(1-v_{i'})^{b}  \,\, \nonumber\\
&  &
\prod_{i'=m'+1}^{s'} v_{i'}^{a}
  (v_{i'}-z)^{c+\kappa}
  (v_{i'}-x)^{-\kappa}(v_{i'}-1)^{b}  \nonumber\\
&  &  \prod_{1\le j<i\le s}(u_i-u_j)^{2/\kappa}
  \prod_{1\le j'<i'\le s'}(v_{i'}-v_{j'})^{2\kappa}
  \prod_{i=1}^s\prod_{i'=1}^{s'}(u_i-v_{i'})^{-2} \,du_i\,dv_{i'}
  \nonumber
\EEA
where $a=2\tilde J_1$, $b=2\tilde J_3$, $c=2\tilde J_2$,
and $d=2(\kappa-1-\tilde J_4)$. Since $a+b+c+d+2-2 \kappa -2 S=0$
the integrals can be parametrized also by $a,b,c$ and $s,s'$.
The above integral is defined for real $z,x\,,$ $0<z<x<1\,.$
The cycle over which the
variables $u_i$, $v_{i'}$ are integrated can be described as
\BA{CC}
  {\cal C}_{m,m'} &=& \{0<u_1<\dots<u_m<z\}\cup
  \{1<u_{m+1}<\dots<u_s<\infty\} \\&&\cup
  \{0<v_1<\dots<v_{m'}<z\}\cup
  \{1<v_{m'+1}<\dots<v_{s'}<\infty\} \,,\nonumber
\EEA
and in this region all powers  in (\ref{II}) can be chosen real,
positive (for real parameters).

Alternatively one can use
\BL{CCC}
  {\cal C}(m,m')
  = \{C_i: i=1,\dots,s\}\cup\{C'_{i'}: i'=1,\dots,s'\}
\EE
where $C^{(')}_{i^{(')}}$ are contours from 0 to $z$
for $i^{(')}=1,\dots,m^{(')}$ and contours from 1 to $\infty$
for $i^{(')}=m^{(')}+1,\dots,s^{(')}$ with $C^{(')}_{i^{(')}+1}$
above $C^{(')}_{i^{(')}}$. Moreover the integrals with
contour ${\cal C}(m,m')$ are taken with an overall phase
\BL{Cphase}
  \exp -{i\pi\over 2}[(m(m-1)+(s-m)(s-m-1))/\kappa
  +(m'(m'-1)+(s'-m')(s'-m'-1))\kappa] \,.
\EE
The contribution from the two alternative cycles differs by a
numerical  factor
\BL{nf}
  \int_{{\cal C}(m,m')} 
  = [m]!\,[s-m]!\,[m']'!\,[s'-m']'!\,   \int_{{\cal C}_{m,m'}}\,,
\EE
where
$$
  [k]=\sin(\pi k/\kappa)/\sin(\pi/\kappa)
  \quad {\rm and} \quad
  [k]'=\sin(\pi k\kappa)/\sin(\pi\kappa)
$$
are  $q=\exp(2i\pi/\kappa)$ and $q'=\exp(2i\pi\kappa)$ numbers
respectively. In the limit $x\to z$
we recover the DF basis of integrals (our notation differs from
that in \cite{DF} by a
shift in the indices $m,m'$and $s,s'$ by one).

The ``even'' integrals in the $t$-basis we will denote by
$\II^{m,m'}$.
The contours of integration are obtained from the above by
interchanging the roles of the points 1 and 0, i.e., we
have $(m,m')$ integrations from $z$
to 1 and $(s-m,s'-m')$ integrations from $-\infty$
to 0. Also the integrand is modified so that again all differences
(when the variables are continued to the  real line) are
positive real numbers and for the $t$-channel we assume that
$0<x<z<1$.

The $u$-channel integrals correspond to contours from 0 to 1
and from $z$ to $\infty$. We will not introduce a special
notation for these integrals.

The ``even'' integrals are not closed under crossing transformations.
If we employ the well known technique of moving contours, e.g., to
express the $s$-channel in terms of the $t$-channel (see the
Appendix for details), because the integrand has a nontrivial
singularity
at $x$ (with respect to the $v_{i'}$ variables) we will generate
also ``odd'' integrals that will involve
contours $\Gamma_x$ ending at, or going around $x$. On the other
hand the singularity at $x$ is like the insertion of a vertex
operator of isospin $j'=1/2\,.$ Because of this if we have one
contour
to $x$
and add a second encircling the previous one the result will be zero.
Thus we have only two possibilities --- no contour to $x$ which
corresponds to the ``even'' integrals and one contour to $x$ which
corresponds to the ``odd'' integrals.

Now we describe a convenient choice  for the ``odd'' integrals of
the $s$-basis which we will denote with
$\FF_{m,m'}$ with $m=0,1,\dots\,, s; m'=1,2,\dots\,, s'\,.$
The  cycle ${}^{(1)}{\cal C}_{m,m'}$ consists of
integrating $v_{i'}$, $i'=1,\dots,m'-1$ from 0 to $z$ while
$v_{m'}$ is integrated along a contour starting at $x$ going around
all the contours from 0 to $z$ in the counterclockwise direction and
returning back to $x$ (i.e., we have a Felder type contour for
$v_{m'}$).
The other integration contours are unchanged.
The integrand is the same as  for $\II_{m,m'}$ except for the
order of the arguments in
the factor $(v_{m'}-z)^{c+\kappa}$. It is convenient
moreover to include in the definition an overall numerical factor
$$
  \FF_{m,m'} =
  { e^{-i\pi(a+c+(2m'-1)\kappa))}
  \over
  \sin(\pi(a+c+(2m'-1)\kappa))}
  \int_{{}^{(1)}{\cal C}_{m,m'}} \dots\,.
$$
The limit $x\rightarrow z$ of the  ``odd'' integrals   $\FF_{m,m'}$
coincides (up to numerical prefactors) with that of the ``even''
ones $\II^{m,m'}$ (for  $m=0,1,\dots\,, s; m'=1,2,\dots\,, s'\,$).

Analogously, the ``odd'' $t$-basis is given by $\FF^{m,m'}$
which differ from the ``even'' ones by trading one integration
from $z$ to 1 for an integration along a Felder type contour
from $x$ around all the contours from $z$ to 1 and
back to $x$ and by appropriate change of the integrand. Also in
the numerical factor above $a$ is changed to $b$.

One can check that the proposed contours correctly correspond to
the relevant basis by taking the asymptotics of the 4-point blocks
$W^{(4)}(x,z)$ as $z,x\to 0$ for the $s$-channel and $z,x\to 1$ for
the $t$-channel and comparing with the corresponding 3-point blocks.
We emphasise that in order to get the correct asymptotics in both
channels we have to take the limit first in $z$ and then in $x$ and
this explains why the $s$-channel integrals we define for $0<z<x<1$
while the $t$-channel ones are defined for $0<x<z<1$.

To find the asymptotics of the ``even'' integrals  one makes a
change of variables $u\to zu$, $v\to zv$ for those variables that
are integrated from 0 to $z$ and takes first $z\to 0$ and then
$x\to 0$ obtaining
\BL{Even}
  \II_{m,m'}(a,b,c,d;z,x) \propto
  z^{-(\Delta_{J_1}+\Delta_{J_2}-\Delta_{J}) -2J_1J_2}
x^{J_1+J_2-J}
  \NN_{m,m'}(a, c+\kappa; \kappa)
  \NN_{s-m,s'-m'}(d,b; \kappa) \,,
\EE
where the intermediate isospin $J$ is given by
\BL{spin}
  m-m'\kappa=J_1+J_2-J
\EE
and $\NN_{m,m'}(\alpha,\beta;\rho)$ are the Selberg integrals
giving the normalization of the 3-point functions which have been
calculated explicitly in \cite{DF}, formula (A.35). In particular
\BL{Se1}
\NN_{m,m'}(a, c+\kappa; \kappa)={\Gamma(1+c-m+m'\kappa)\,
\Gamma(2+a+c-m+(m-1)'\kappa) \over
\Gamma(1+c)\,\Gamma(2+a+c-2m+(2m'-1)\kappa)   }\,
 \NN_{m,m'}(a, c; \kappa) \,.
\EE
Recall that the chiral correlators (\ref{scor})  differ from the
integrals by the overall
factor $ z^{2J_1J_2}(1-z)^{2J_2J_3}\,,$ which will compensate
 $-2J_1J_2$ in the power of $z$ above, and thus producing exactly
the functional dependence of a 3-point function.

The asymptotics of the ``odd'' integrals is performed as
follows. With the integrals from 0 to $z$ proceed as above.
The integral over $v_{m'}$ express as a combination of two
integrals from $0$ to $x$, one going above, the other below $z$,
and in each of them make the change of variables
$v_{m'}\to xv_{m'}$ and again take first $z\to 0$ and then $x\to
0$. The result is
\BA{Odd}
  \FF_{m,m'}(a,b,c,d;z,x) & \propto &
  z^{-(\Delta_{J_1}+\Delta_{J_2}-\Delta_{J}^{(1)}) -2J_1J_2} \,
  x^{J_1+J_2-
  J^{(1)}} \qquad\qquad\qquad\qquad \nonumber \\
& &
\NN_{0,1}(2J-\kappa,-\kappa;\kappa)
  \NN_{m,m'-1}(a, c+\kappa;\kappa)
  \NN_{s-m,s'-m'}(d,b;\kappa) \,,
\EEA
where $J^{(1)}= \kappa -1-J  $ with $J$ as in (\ref{spin})
and $\Delta_J^{(1)}\equiv J^{(1)}(J^{(1)}+1)/\kappa =
\Delta_J+\kappa-2J-1$. Note that the prefactor we had included
in the definition of $\FF_{m,m'}$ has cancelled on the r.h.s.
above.
 The normalisation constants can be again expressed through the
ones in \cite{DF}
\BA{Se2}
&&\NN_{0,1}(2J-\kappa,-\kappa;\kappa)\, \NN_{m,m'-1}(a, c+\kappa;
\kappa)  \nonumber \\
&&={\Gamma(\kappa)\, \Gamma(1-\kappa)\,
\Gamma(1+a+c-2m+(2m-1)'\kappa) \over
\Gamma(1+c)\,\Gamma(-m+m'\kappa)\,\Gamma(1+a-m+(m'-1)\kappa)
}\,
 \NN_{m,m'}(a, c; \kappa) \,. 
\EEA

We conclude that
the asymptotics of the ``even'' and ``odd'' integrals is
qualitatively in agreement with the  fusion rules (\ref{frs}).

For the asymptotics of the $t$-basis integrals
analogously first we take $z\to 1$ and then $x\to 1$.
Again one checks that the identification between choice of
contours and intermediate isospins is correct.

Note that the above choices of ``odd'' integrals is not unique.
Instead of the Felder type contours one could also take e.g., an
arbitrary combination of the constituent contours going from $0$
to $x$, above or below $z\,,$ the corresponding integrals to be
denoted $  \FF_{m,m'}^{+}$ and $  \FF_{m,m'}^{-}$ respectively.
Furthermore there is a
certain ambiguity
in the way  the limit ``first $z$, then $x$ goes to $0$'', is
taken.

\sect{Braid and fusion transformations,
local correlators and structure constants}
%
%

The duality transformations describing the exchange (braiding) of
several punctures or the connection between the different possible
sewings of the $n$-punctured sphere out of 3-punctured spheres
translate into matrix transformations acting on the basis of
conformal blocks introduced in the previous section. The physical
correlators are built as quadratic forms of left and right chiral
blocks. The principle of locality (i.e., the symmetry of the
euclidean functions) demands that these quadratic
forms are invariant under the duality transformations. Once the
local two dimensional 4-point functions have been determined and
the normalizations of the 3-point functions have been fixed one
can easily read off the operator product coefficients.

It is important to emphasise that our sets of 4-point blocks
realise a representation of the braid group on 4-strands, each
strand or puncture corresponding to a pair $(x_a,z_a)$,
$a=1,\dots,4$. Thus we are moving each pair as a whole, e.g.,
if we consider the monodromy of the second puncture with respect
to the first we should move $z_2$ and $x_2$ simultaneously
around the pair $(x_1,z_1)$ along homotopic loops (equivalently
we should move the ratios $z$ and
$x$ simultaneously around 0 along homotopic loops).

To describe the duality matrices it is enough to specify the
diagonal braidings in the different channels and the crossing
(fusion) matrices between two channels.

For example the diagonal braiding in the $s$-channel is the
exchange of the first two punctures. The result is
$$
  W_J(z_1,x_1,J_1; z_2,x_2,J_2; z_3,x_3,J_3; z_4,x_4,J_4)
  \qquad\qquad\qquad\qquad
$$
$$
  \qquad\qquad\qquad\qquad
  = \exp(\pm i\pi(h_{J_1}+h_{J_2}-h_J)) \,
  W_J(z_2,x_2,J_2; z_1,x_1,J_1; z_3,x_3,J_3; z_4,x_4,J_4) \,.
$$
Here $W_J$ are the full chiral correlators, including the
prefactors. The ($-$)$+$ sign corresponds to (counter) clockwise
exchange of $(x_1, z_1)$ and  $(x_2, z_2)$, and $h_J$ are the
``reduced'' conformal weights (\ref{h(J)-J}),  $h_J=\Delta_J-J$. As
already pointed out the weights $h_J$ are invariant under the
transformation $J\to\kappa-1-J\,,$ hence the ``even'' and ``odd''
blocks have the same diagonal braidings.
The appearance of the ``reduced'' conformal weights and not the
Sugawara weights $\Delta$ in the phases
is due to the fact that the
analytic continuation of the integrals is done moving the pairs
$(x_a,z_a)$
as a whole.

Now we describe the crossing matrices expressing the $s$-basis in
terms of the $t$-basis. All details and explicit expressions are
left for the appendix. The standard $z$-dependent prefactor relating
the blocks and the integrals plays no role in the crossing matrices
so we define them in terms of the integrals introduced in the
previous section. It is convenient first to introduce a new basis of
integrals in which the fusion transformation takes a block diagonal
form. Let
\begin{eqnarray}
\YY_{mm'} &=& h_{m'} \II_{mm'} - f\,\, \FF_{mm'} \nonumber\\ &&\\
\ZZ_{mm'} &=& g_{m'} \II_{mm'} +   \FF_{mm'} \nonumber
\end{eqnarray}
for $m'=1,2,\dots \,, s'\,,$
while $\YY_{m\,0}\equiv \II_{m\,0}\,,$ where
\BL{hh}
  h_m = {\sin(\pi(a+c+(m-1)\kappa))
         \sin(\pi(c+m\kappa)) \over
                 \sin(\pi(a+c+(2m-1)\kappa))
                 \sin(\pi c)}\,, \quad
  f =   {\sin(\pi\kappa)\over \sin(\pi c)}\,,
\EE
\BL{ffgg}
  g_m  = -{[m]' \sin(\pi(a+(m-1)\kappa))
            \over
                  \sin(\pi(a+c+(2m-1)\kappa))
                  }= {1-h_m\over f}\,.
\EE

The inverse transformation is
\begin{eqnarray}
\II_{mm'} &=& \YY_{mm'} + f\,\, \ZZ_{mm'} \nonumber\\&&\\
\FF_{mm'} &=& -g_{m'}\YY_{mm'} + h_{m'}\, \ZZ_{mm'}\,. \nonumber
\end{eqnarray}
For the $t$-channel analogs
(the integrals with upper indices $m,m'$)
we have analogous relations with $h^m$ and $g^m$ being the same as
$h_m$ and $g_m$ only $a$ is substituted with $b$.
The integrals $\ZZ$  are described by cycles
in which instead of the
Felder type countour starting and ending at $x$ we have  simply a
contour from $z$ to $x$ and all the rest is as in $\FF$.

In the $\{\YY,\ZZ\}$ basis the fusion transformation
 takes a particularly simple form
\BA{fusm}
\YY_{mm'} &=& \sum_{nn'} \alpha_{mm',nn'}\,\YY^{nn'}\nonumber\\&&\\
\ZZ_{mm'} &=& \sum_{nn'} \gamma_{mm',nn'}\,\ZZ^{nn'}\,,\nonumber
\EEA
where
\BA{alga}
  \alpha_{mm',nn'} &=&
  \alpha_{mn}(-a/\kappa,-b/\kappa,
              -c/\kappa,-d/\kappa;1/\kappa)\, \,
  \alpha_{m'n'}(a,b,c,d;\kappa)
\\
  \gamma_{mm',nn'} &=& \theta\,
  \alpha_{mn}(-a/\kappa,-b/\kappa,
              -c/\kappa,-d/\kappa;1/\kappa)\, \,
  \alpha_{m'-1,n'-1}
         (a,b,c+2\kappa,d;\kappa)
\EEA
The matrices $\alpha$ coincide with the crossing matrices between
the $s$- and $t$-channels in the Virasoro minimal models.
The derivation of (\ref{fusm}) is outlined in the Appendix.
Note that the second factor in $\gamma$ differs from that in $\alpha$
by a simple shift of the parameters. The phase
$\theta=\exp i\pi(c+1)$ appears because transforming the $s$-channel
``odd'' integral into $t$-channel ``odd'' ones we have also to
``braid'' $x$ and $z\,,$ exchanging their places. The
properly normalized matrices $\alpha$ are given up to signs by a
product of  $q -$ and $q'-$ $6j$-symbols.
Note that these symbols (see \cite{KR} for an explicit
expression) are invariant (up to sign) under the
change  $j_1 \rightarrow {1\over 2} (j_1+j_2-j_3-j_4-1)$, etc.,
implied by the ``tilde'' transformation (\ref{tJ}); the multiples
of $\kappa$ lead to an overall sign.
To see this one has to use at an intermediate step the property
$$
\left\{\begin{array}{c} j_1\ j_2\ j_5 \\   j_3 \ \overline{j}_4 \
j_6 \end{array}   \right\}_q = (-1)^{2j_4+1} \
\left\{\begin{array}{c} j_1\ j_2 \ j_5 \\   j_3 \  j_4  \  j_6
\end{array}\right\}_q \ , \qquad    \overline{j}_4=-1-j_4 \,, $$
derived in \cite{GP89}.

In the limit $x\to z$ the elements $\YY $ of the
basis, ``block diagonalising'' the  fusion transformations,
reduce to the  corresponding DF integrals while $\ZZ$ map to zero.

Now we discuss the possible 2-dimensional local 4-point correlators.
They are invariant with respect to braid and fusion transformations
quadratic forms of the chiral blocks.
Since according to (\ref{fusm}) the
subset of integrals $\YY$ behave under braid and fusion
transformations as the minimal model DF integrals we immediately
have (omitting the $z$ dependent prefactors) an invariant
quadratic form
\BL{newi}
  \sum_{m=0}^{s}\sum_{m'=0}^{s'} X_{mm'}\, |\YY_{mm'}|^2
\EE
where
$X_{mm'}=X_{m+1}^{(s+1)}(-a/\kappa,-b/\kappa,-c/\kappa
;1/\kappa)
\,X_{m'+1}^{(s'+1)}(a,b,c;\kappa)$ are the same as the ones
found in \cite{DF}
\BA{X}
   X_{m+1}^{(s+1)}
   (a,b,c
   ;\rho) &=&
 [m]!\,[s-m]!\,
 \prod_{k=0}^{m-1}
        {\sin(\pi(1+a+k\rho)) \, \sin(\pi(1+c+k\rho))  \over
          \sin(\pi(2+a+c+(m-1+k)\rho))}              \nonumber \\
&& \prod_{k=0}^{s-m-1}
        {\sin(\pi(1+b+k\rho)) \,
                 \sin(\pi(1+d+k\rho))  \over
         \sin(\pi(2+b+d+(s-m-1+k)\rho))}\,.
\EEA
Since the crossing matrices $\gamma$ for the ``odd'' integrals
$\ZZ$ are obtained from the ``even'' crossing matrices $\alpha$
by a certain shift of the parameters the ``odd'' invariant
\BL{sinv}
  \sum_{m=0}^{s}\sum_{m'=1}^{s'} {}^{(1)}X_{mm'}\, |\ZZ_{mm'}|^2
\EE
is obtained by the same shift. From the explicit form of $X_{mm'}$
we have
\BL{Xe}
  {}^{(1)}X_{mm'} =- {h_{m'}
  \over  g_{m'} \sin(\pi(c+\kappa))} X_{mm'}\,,
\EE
where $h_m$ and $g_m$ were defined in (\ref{hh}), (\ref{ffgg}).

Having two independent invariants we obtain a 1-parameter family
of monodromy invariants
\BL{fam}
  M(\xi) \propto \sum_{mm'} X_{mm'} \left( |\YY_{mm'}|^2
  + \xi {fh_{m'}\over g_{m'}} |\ZZ_{mm'}|^2 \right)\,.
\EE
Going back to the basis of $\II$ integrals this invariant
is nondiagonal except for $\xi=1$ in which case we get
\BA{Andi}
 M(1) &\propto & \sum_{mm'} (h_{m'}\,X_{mm'})\, |\II_{mm'}|^2 +
  \sum_{mm'} ({f\, \over g_{m'}}\,X_{mm'})\, |\FF_{mm'}|^2 \\
&=&
 \sum_{mm'}   \, {\sin(\pi(2+a+c-m+(m'-1)\kappa))
         \sin(\pi(1+c-m+m'\kappa)) \over
                 \sin(\pi(2+a+c-2m+(2m'-1)\kappa))
                 \sin(\pi (1+c))}\,\,X_{mm'}\, |\II_{mm'}|^2 \nonumber \\
& +&
 \sum_{mm'} \,{\sin^2(\pi\kappa) \,
\sin(\pi(1+a+c-2m+(2m'-1)\kappa))
                    \over \sin(\pi(-m+m'\kappa))\, \sin(\pi
(1+c))\,\sin(\pi(1+a+(m'-1)\kappa))} \,X_{mm'}
\, |\FF_{mm'}|^2   \,. \nonumber
\EEA
(The summation over $m'$ in the second terms in (\ref{fam}) and
(\ref{Andi}) starts from $m'=1$.) The numerical coefficients in the
$t$-channel  ($u$-channel) are obtained exchanging $a
\leftrightarrow b$ ($a \leftrightarrow d$) respectively.

A straightforward though rather lengthy computation, which we omit,
shows that the invariant $M(1)$ coincides precisely, up to the
standard prefactors, with the (explicitly local) volume integral
for the 4-point correlator  \cite{And};
the derivation of this result is based on the technique
explained in \cite{Dot}. The choice of ``odd'' integrals
$\FF$ as described above is thus selected by  their property to
diagonalise the invariant of \cite{And} and this is the way we have
originally arrived at it.

The expression (\ref{Andi})   is still formal
since in general the summation may include terms violating
the fusion rules (\ref{frs}). The cancellation of the unwanted
terms amounts
in general to some linear relations for the set of integrals (see
\cite{FGP1} for the analogous consideration in the Virasoro case);
we shall not address this problem here.

{}From (\ref{Andi}) one can read the
product $D_{J_1 J_2}^{J_5^{(t)}}\, D_{J_5^{(t)} J_3}
^{J_4}\,$ of  OPE structure constants. Here as above $t =0$, or
$1$, corresponding to the intermediate isospin $J_5=J_5^{(0)}$ or
$J_5^{(1)}$ respectively. (To simplify the notation we write
$J_i$ instead of the pair $(J_i, J_i)$ of left-right isospins  in
these 2-dimensional constants.)   We shall assume
that the even constants are normalised according to
 $D_{J J}^{0}=1 = D_{J 0}^{J}$. Normalizing
 the two point functions to one together with the locality
of the 3-point correlators imply that the (even) structure
constants are symmetric with respect to all indices;
in particular $D_{J_1 J_2}^{J_5} = D_{J_1 J_5}^{J_2}\,$. To
recover the constants one has to compare the $s$- and
$t$-channels of (\ref{Andi}) in the case of two by two coinciding
fields $J_1=J_4\,, J_2=J_3$ -- hence $a=\kappa-1\,,$ $ c=1-\kappa
+2J_1+2J_2\,,$ $ b=\kappa-1 +2J_2-2J_1\,,$ $S=2J_2\,.$ Then the
identity operator appears  in the $t$-channel and one multiplies
the correlator by an overall constant determined by the
normalisation requirement  $D_{J_2 J_2}^{0}=1=D_{J_1 0}^{J_1}\,,$
i.e., this constant is given by the inverse of the square of the
normalisation constant in
(\ref{Even})   times the constant in front of the
first term in  (\ref{Andi}), both taken for $m=s\,, m'=s'\,,$ and
with $a\leftrightarrow b$. This determines the structure constants
for any intermediate isospin. Taking into account
 (\ref{Even}), (\ref{Se1}),  (\ref{Odd}), (\ref{Se2}),
(\ref{Andi}) one obtains
\BL{strc}
  D_{J_1 J_2}^{J_5^{(t)}}\,D_{J_1 J_5^{(t)}}^{J_2}
  ={P^2 (2-\kappa+J_1+J_2+J_5^{(t)}) \, P(2-\kappa)\over
  P(2-\kappa +2J_1)\, P(2-\kappa +2J_2)\,
  P(2-\kappa +2J_5^{(t)})\,}
  \, (D_{J_1 J_2}^{J_5;DF})^2\,,
\EE
where $P(a):=\Gamma(a)/\Gamma(1-a)$. The structure constant
$D_{J_1 J_2}^{J_5;DF}$ in
the r.h.s. is precisely the corresponding DF constant, see
\cite{DF3}.
 The r.h.s. of (\ref{strc}) has sense also
for the values
of $J_a$ beyond the fusion table (and the fusion rules) of the
minimal models;
 in particular the zeros of the DF constant are
compensated by singularities of the prefactor.
{}From (\ref{strc}) it is clear that we can also require for the
odd constants the symmetry properties mentioned above, and hence
replace the l.h.s. with the square of the constant
$D_{J_1 J_2}^{J_5^{(t)}}$; in other words,  we can identify in
averages the physical operators
$\Psi_{(J_{r,r'}^{(1)}\,,\,J_{r,r'}^{(1)})}^{(1)}\,$  and
$\Psi_{(J_{p-r-2, p'-r'-2}\,,\,J_{p-r-2, p'-r'-2)}}^{(0)}\,,$ cf.
(\ref{afW}). In the even case $ t =0$
the above expression (which can be looked as the
analytic continuation of the expression for the
constants of Fateev and Zamolodchikov \cite{ZF} in the
case  of (half)integer isospins) first appeared  in \cite{And}.
The constants $D_{J_1 J_2}^{J_5^{(t)}}$ are not real in general.

Another important special case of the invariants (\ref{fam})
is provided by  $M(0)$, i.e., (\ref{newi}), which allows furthermore
for a ``heterotic''  interpretation having $A^{(1)}_1$ as chiral
algebra for the left movers and pure Virasoro for the right movers.
Indeed the braid and monodromy properties of the integrals
$Y_{m m'}(x,z)$ imply that they can be  coupled with the
corresponding
(complex conjugate) DF integrals $I_{m m'}^{(\rm{DF})}(z)$ to produce
an invariant analogous to (\ref{newi}). The resulting structure
constants can be computed as above. Such local correlators might
be of relevance for the program outlined in \cite{KPZ}.

We conclude with a few remarks. One has to check that the
obtained crossing matrices satisfy the polynomial (in
particular the pentagon) equations determined by the Awata-Yamada
fusion rules. This in particular could fix the ambiguity in the
physical interpretation of the ``odd'' correlators. As mentioned
at the end of Section 4,  instead
of   $\FF_{m,m'}$ we could interpret $  \FF_{m,m'}^{+}$ or
$  \FF_{m,m'}^{-}$ as
``odd'' correlators.  The set $\{\II_{m,m'} \,,\,
\FF_{m,m'}^{+}\}$ transforms via a fusion matrix whose
``even - even'' block has a lot of vanishing
matrix elements. E.g., for coinciding isospins
$J_1=J_2=J_3=J_4$ and $J_5 < J_1+J_2$ there is no contribution of
``even'' isospins
$J_6$ apart from  $J_6=J_2+J_3$ -- this behaviour is partially in
agreement with the observation in the early paper \cite{Do} --
only unlike  \cite{Do} we have in addition nonvanishing fusion
matrix  elements corresponding to the  ``odd''
isospins.\footnote{ Note that this property
of the fusion matrix in the basis  $\{\II_{m,m'} \,,\,
\FF_{m,m'}^{+}\}$ still does not contradict the fusion rules
(\ref{frs}), since there are also nonvanishing  matrix elements
corresponding to the admissible triples
$(J_2\,, J_3\,, J_6)$ with $J_6 \le J_2+J_3$.
The phenomena of vanishing of some fusion matrix elements for
triples consistent with the fusion rules is known to occur
already in the Virasoro case.}

The supersymmetric interpretation of the fusion rules
\cite{FM} raises also the question whether
the crossing matrices could be related to the $6j$-symbols of
the $q$-analog of the relevant superalgebra.

\bigskip
{\bf Acknowledgements}

It is a pleasure to thank Vl. Dotsenko, F. Malikov,  V. Schomerus
and A. Varchenko for useful discussions at different stages of
this work. V.B.P. acknowledges  the support of INFN, Trieste, and
A.Ch.G. and V.B.P.  acknowledge
the support of Erwin Schroedinger Institute (ESI), Vienna.
This work  has been also partially  supported by the Bulgarian
Foundation for Fundamental Research under Contract $\phi$-404-95.

\bigskip\bigskip

Note added:

While this paper was being written down\footnote{A preliminary
version has  been reported by one of us (A.Ch.G.) in May, 1996,
at the program on Topological, Conformal and Integrable Quantum
Field Theory of ESI,  Vienna} a paper
by J.L. Petersen, J. Rasmussen and M. Yu \cite{PRYn},
has appeared,  the results of which overlap partially with ours.

\newpage
\renewcommand{\theequation} {A.\arabic{equation}}
\appendix
\sect{Appendix}
%
%

In this appendix we will derive the formulae for the crossing
matrices. As in \cite{DF} the fusion transformation in the general
case $s\not =0\, s'\not =0$ factorises due to the trivial
monodromy of the mixing factor
$\prod\, (u_i-v_{i'})^{-2}\,$ in (\ref{PhiPhi}) into two
pieces. The derivation of the first repeats the one relevant for
the correlators in the Virasoro models
because of the trivial monodromy of the 
modifying factors,
$\prod_{i=1}^s\, (u_i-z)^{-1}\,(u_i-x) \,,$ in (\ref{sphi}).
So we will concentrate
on the  derivation of the second piece assuming furthermore
for simplicity of notation that  $s=0\,.$
In order to make the formulae more readable we will
skip the primes that we have consistently put on all variables
related to this case until now.

Define a generalization of the ``even'' integrals $\II$ in section 4
$$
  \II[m_1,m_2,m_3,m_4] = \int_{{\cal C}(m_1,m_2,m_3,m_4)}\dots
$$
Here the contour is a generalization of the contour (\ref{CCC})
(including an analog of the phase (\ref{Cphase})) in which there
are $m_1$ contours from $-\infty$ to 0, $m_2$ --- from 0 to $z$,
$m_3$ --- from $z$ to 1, and $m_4$ from 1 to $\infty$.
A word of caution --- these integrals we will need only when either
$m_2=0$ or $m_3=0$ and our convention (in order not to overburden
the notation) is that if $m_2=0$ then the integral is defined for
$0<x<z<1$ while if $m_3=0$ then the integral is defined for
$0<z<x<1$. Note that the integral $\II[m_1,0,0,m_2]$ is well
defined (and identical)
in both cases because there is no relative monodromy between the
points $z$ and $x$.
Up to the $q$-factorials in (\ref{nf}) we have that
$\II[0,m,0,s-m]$ is equal to $\II_{0m}$ and $\II[s-m,0,m,0]$ is
equal to $\II^{0m}$.

Now we define ``odd'' integrals $\ZZZ$ that are
generalizations of the $\ZZ$ ones.
For $\ZZZ[m_1,0,m_3,m_4]$ we assume that $0<x<z<1$ and that there
are $m_1$ contours from $-\infty$ to 0, one contour from $x$ to
$z$, $(m_3-1)$ contours from $z$ to 1, and $m_4$ contours from 1
to $\infty$.
For $\ZZZ[m_1,m_2,0,m_4]$ we assume that $0<z<x<1$ and that there
are $m_1$ contours from $-\infty$ to 0,  $(m_2-1)$ contours from
0 to $z$, one contour from $z$ to $x$, and $m_4$ contours from 1
to $\infty$.
Up to the $q$-factorial prefactors we have that
$\ZZZ[0,m,0,s-m]$ is equal to $\ZZ_{0m}$ while
$\ZZZ[s-m,0,m,0]$ is equal to $\ZZ^{0m}$.
Now the monodromy between $x$ and $z$ is nontrivial and we have
\BL{theta}
\ZZZ[m_1,1,0,m_4] = \theta \ZZZ[m_1,0,1,m_4]
\EE
where $\theta=\exp(\pm i\pi (c+1))\,$ depending on whether
we exhange $x$ and $z$ in a clockwise or counterclockwise direction.

The transformation of $s$-channel integrals (the ones with
subscripts) into $t$-channel integrals (the ones with
superscripts) is performed in two steps. The first step consists
in reducing $m_2$ keeping $m_3=0$ fixed until we get $m_2=0$
(or $m_2=1$ for the ``odd'' integrals). The second step consists
in reducing $m_4$ keeping $m_2=0$ fixed.

In order to obtain recursion formulae allowing us to move
contours from one interval to another one adds a closed contour
above/below  the real axis and substracting the resulting two
equations with appropriate coefficients one gets the relations
\BA{II2}
&&\sin(\pi(a+c+(m_1+2m_2)\kappa))\, \II[m_1+1,m_2,0,m_4] +
\sin(\pi(c+m_2\kappa))\, \II[m_1,m_2+1,0,m_4]  \nonumber\\
&&- \sin(\pi(b+m_4\kappa))\, \II[m_1,m_2,0,m_4+1]
  - \sin((\pi\kappa))\, \ZZZ[m_1,m_2+1,0,m_4]=0 \,,
\EEA
\BA{ZZZ2}
&&\sin(\pi(a+c+(m_1+2m_2)\kappa))\, \ZZZ[m_1+1,m_2,0,m_4]
+\sin(\pi(c+(1+m_2)\kappa))\, \ZZZ[m_1,m_2+1,0,m_4]  \nonumber\\
&&-\sin(\pi(b+m_4\kappa))\, \ZZZ[m_1,m_2,0,m_4+1] =0 \,,
\EEA
which are relevant for the first step and the relations
\BA{II4}
&&\sin(\pi(a+m_1\kappa)) \,\II[m_1+1,0,m_3,m_4] -
\sin(\pi(c+m_3\kappa))\, \II[m_1,0,m_3+1,m_4] \\ \nonumber
&&-\sin(\pi(b+c+(2m_3+m_4)\kappa))\, \II[m_1,0,m_3,m_4+1]
+\sin((\pi\kappa))\,\ZZZ[m_1,0,m_3+1,m_4]=0 \,,
\EEA
\BA{ZZZ4}
&&\sin(\pi(a+m_1\kappa))\, \ZZZ[m_1+1,0,m_3,m_4] -
\sin(\pi(c+(1+m_3)\kappa))\, \ZZZ[m_1,0,m_3+1,m_4]  \nonumber\\
&&-\sin(\pi(b+c+(2m_3+m_4)\kappa))\, \ZZZ[m_1,0,m_3,m_4+1] =0\,,
\EEA
which are relevant for the second step. Note that the
corresponding recursion relations for the fusion transformation
of the DF integrals are recovered from (\ref{II2}) and (\ref{II4})
 simply dropping the last terms.

Next define coefficients $A$, $B$, and $C$ by
\BA{}
    \II[m_1,m_2,0,m_4] &=& \sum\,
    A^{(2)}_{k_1,k_4} \,
    \II[m_1+k_1,m_2-k_1-k_4,0,m_4+k_4]
\nonumber \\
  &+&  \sum\, B^{(2)}_{k_1,k_4}\,
    \ZZZ[m_1+k_1,m_2-k_1-k_4,0,m_4+k_4]
        \,, \nonumber
\\
  \ZZZ[m_1,m_2,0,m_4] &=& \sum
  C^{(2)}_{k_1,k_4}\,
  \ZZZ[m_1+k_1,m_2-k_1-k_4,0,m_4+k_4]\,, \nonumber
\EEA
relevant for the first step
(the sums are over $k_1$ and $k_4$ with $k_1+k_4$ fixed)
and
\BA{}
  \II[m_1,0,m_3,m_4] &=& \sum\,
  A^{(4)}_{k_1,k_3}\,
  \II[m_1+k_1,0,m_3+k_3,m_4-k_1-k_3]
\nonumber \\
&+&  \sum\, B^{(4)}_{k_1,k_3}\,
  \ZZZ[m_1+k_1,0,m_3+k_3,m_4-k_1-k_3]
  \,,
\nonumber \\
  \ZZZ[m_1,0,m_3,m_4] &=& \sum
  C^{(4)}_{k_1,k_3}\,
  \ZZZ[m_1+k_1,0,m_3+k_3,m_4-k_1-k_3]\,,
\nonumber\EEA
relevant for the second step.
The relations (\ref{II2}), (\ref{ZZZ2}) and (\ref{II4}),
(\ref{ZZZ4}) translate into recursion relations for the
coefficients $A$, $B$, $C$
\BA{AB2R} \nonumber
 - A^{(2)}_{k_1k_4} &=&
  { \sin(\pi(-a{-}c{+}(k_1{+}2k_4{-}m_1{-}2m_2{+}1)\kappa))
   A^{(2)}_{k_1-1\,k_4}
   + \sin(\pi(b{+}(m_4{+}k_4{-}1)\kappa)) A^{(2)}_{k_1\,k_4-1}
   \over
   \sin(\pi(-c+(k_1+k_4-m_2)\kappa))} \\
  - B^{(2)}_{k_1k_4} &=&
   {\sin((\pi\kappa)) A^{(2)}_{k_1k_4}
  \over \sin(\pi(-c+(k_1+k_4-m_2+1)\kappa))}
\\ \nonumber &+&
  {  \sin(\pi(-a{-}c{+}(k_1{+}2k_4{-}m_1{-}2m_2{+}1)\kappa))
  B^{(2)}_{k_1-1\,k_4}
   +  \sin(\pi(b{+}(m_4{+}k_4{-}1)\kappa)) B^{(2)}_{k_1\,k_4-1}
   \over
   \sin(\pi(-c+(k_1+k_4-m_2-1)\kappa))}
\EEA
and
\BA{AB4R} \nonumber
 A^{(4)}_{k_1,k_3} &=&
   { \sin(\pi(a+(m_1+k_1-1)\kappa)) \,A^{(4)}_{k_1-1,k_3}
   \over \sin(\pi(b+c+(2m_3+m_4+k_3-k_1)\kappa))}
  -
  {\sin(\pi(c+(m_3+k_3-1)\kappa))\,A^{(4)}_{k_1,k_3-1}
     \over \sin(\pi(b+c+(2m_3+m_4+k_3-k_1-2)\kappa))}
\\
  B^{(4)}_{k_1,k_3} &=&
   {\sin(\pi(a+(m_1+k_1-1)\kappa))\,B^{(4)}_{k_1-1,k_3}
   \over  \sin(\pi(b+c+(2m_3+m_4+k_3-k_1)\kappa))}
\\ \nonumber &+&
  {-\sin(\pi(c+(m_3+k_3)\kappa))\,B^{(4)}_{k_1,k_3-1}
  + \sin((\pi\kappa)) A^{(4)}_{k_1,k_3-1}
   \over  \sin(\pi(b+c+(2m_3+m_4+k_3-k_1-2)\kappa))}
\EEA
The solutions of (\ref{AB2R}) are
\BA{AB2}
  A^{(2)}_{k_1,k_4}
  &=& (-1)^{k_1+k_4}
  {[k_1+k_4]!\over[k_1]![k_4]!}  \\ && \nonumber
  { \prod_{i=0}^{k_1-1}\sin(\pi(-a-c+(2-m_1-2m_2+k_4+i)\kappa))
  \prod_{i=0}^{k_4-1}\sin(\pi(b+(m_4+i)\kappa))
  \over \prod_{i=0}^{k_1+k_4-1}\sin(\pi(-c+(1-m_2+i)\kappa))}\,,
\\
  B^{(2)}_{k_1,k_4}
  &=& {-[k_1+k_4+1]\sin((\pi\kappa))\sin(\pi(-c+(k_1+k_4-m_2)\kappa))
  \over \sin(\pi(-c-m_2\kappa))\sin(\pi(-c+(k_1+k_4-m_2+1)\kappa))}\,
  A^{(2)}_{k_1,k_4}\,,
\EEA
while the solutions of (\ref{AB4R}) are
\BA{AB4}
  A^{(4)}_{k_1,k_3}
  &=&
  {(-1)^{k_3} [k_1+k_3]!\over[k_1]![k_3]!}\,
    {\prod_{i=0}^{k_1-1}\sin(\pi(a+(m_1+i)\kappa))
  \over
  \prod_{i=0}^{k_1-1}\sin(\pi(b+c+(2m_3+m_4+k_3-k_1+i)\kappa))}  \\
\nonumber
  && {\prod_{i=0}^{k_3-1}\sin(\pi(c+(m_3+i)\kappa))
  \over
  \prod_{i=0}^{k_3-1}\sin(\pi(b+c+(2m_3+m_4-k_1-1+i)\kappa))}\,,
\\
  B^{(4)}_{k_1,k_3}
  &=& {-[k_3]\sin((\pi\kappa))\over
  \sin(\pi(c+m_3\kappa))}\,  A^{(4)}_{k_1,k_3}\,.
\EEA
The recursion relations for $C$ are the same as for $A$ except
for a shift of the parameters and their solutions are
\BA{C24}
  C^{(2)}_{k_1,k_4} &=& {\sin(\pi(c+(m_2-k_1-k_4)\kappa))
  \over \sin(\pi(c+m_2\kappa))}\,
  A^{(2)}_{k_1,k_4}\,,
\\
  C^{(4)}_{k_1,k_3} &=& {\sin(\pi(c+(m_3+k_3)\kappa))
  \over \sin(\pi(c+m_3\kappa))}\,
  A^{(4)}_{k_1,k_3}\,.
\EEA

Putting the coefficients $A,B,C$ from the two steps together
it is immediate to write the crossing matrices in a mixed basis
consisting of $\II$'s in the even part and $\ZZ$'s in the odd part.
In this basis the fusion transformation takes a
block triangular form
$\II_{mm'}=\sum\alpha_{mm',nn'} \II^{nn'} +
\beta_{mm',nn'} \,\ZZ^{nn'}$,
$\ZZ_{mm'}=\sum\gamma_{mm',nn'} \,\ZZ^{nn'}$.
Schematically we have
$\alpha = A^{(2)} * A^{(4)}$,
$\gamma = C^{(2)} * \theta C^{(4)}$,
$\beta = A^{(2)} * B^{(4)} + B^{(2)} * \theta C^{(4)}$.
For example for  $\alpha_{m,n}$ using the above solutions for
$A^{(2)}$ and $A^{(4)}$ and remembering that we have to include
the factorials from (\ref{nf}) we obtain
\BA{alpha}
  \alpha_{mn} &=&
  {[s-n]!\over[s-m]!}
  {1 \over \prod_{i=0}^{m-1} \sin(\pi(-c-i\kappa)}
  \prod_{i=0}^{n-1} {\sin(\pi(c+i\kappa))
                    \over \sin(\pi(b+c+(n-1+i)\kappa))} \,
 \nonumber \\ &&
  \sum_k {[k]! \over [s-k]![k-n]![k+m-s]!}
  \prod_{i=0}^{s-k-1}  \sin(\pi(-a-c+(1-m-i)\kappa)) \\&&
  \prod_{i=0}^{k+m-s-1} \sin(\pi(b+(s-m+i)\kappa))   \nonumber
  \prod_{i=0}^{k-n-1}  { \sin(\pi(a+(s-k+i)\kappa))
                         \over \sin(\pi(b+c+(2n+i)\kappa))}  \,.
\EEA
This  expression,   properly normalised by the square root of the
ratio $X_{m+1}^{s+1}(a,b,c)/X_{n+1}^{s+1}(b,a,c)\,,$ cf. (\ref{X}),
can be brought (modulo signs) into the standard Racah form of the
$6j$-symbols, see
\cite{lit} for the classical counterparts of the various
representations of these symbols.

The $\alpha$'s and $\gamma$'s here are the same as  the ones in
the $\{\YY,\ZZ\}$ basis. It is a simple algebra to obtain the
crossing matrices in the $\{\II,\FF\}$ basis.

%
%


\newpage
\begin{thebibliography}{99}
   \newcommand{\bi}{\bibitem} \newcommand{\wb}{\linebreak[0]}

\bi{KZ}
   V.G. Knizhnik and A.B. Zamolodchikov, {\it Nucl. Phys.}
   {\bf B247} (1984) 83.
\bi{SV}
   V.V. Schechtman and A.N. Varchenko,
   {\it Lett. Math. Phys.} {\bf 20} (1990) 279;
   {\it Inven. Math.} {\bf 106} (1991) 139.
\bi{ZF}
   A.B. Zamolodchikov and V.A. Fateev,
   {\it Sov. J. Nucl. Phys.} {\bf 43} (1986) 657.
\bi{ChF}
   P. Christe and R. Flume, {\it Nucl. Phys.} {\bf B282}
   (1987) 466,\\
   P. Christe, PhD Thesis, Bonn University (1986).
\bi{DJMM}
   E. Date, M. Jimbo, A. Matsuo and T. Miwa, in: Yang-Baxter
   Equations, Conformal Invariance and Integrability in
   Statistical Mechanics and Field Theory, World Scientific (1989).
\bi{KW}
   V.G. Kac and M. Wakimoto, {\it Proc. Natl. Acad. Sci. USA}
   {\bf 85} (1988) 4956;\\
   V.G. Kac and M. Wakimoto, {\it Acta Appl. Math.}
   {\bf 21} (1990) 3.
\bi{Gaw} K. Gawedzki, manuscript on CFT, in preparation.
\bi{BO}
   M. Bershadsky and H. Ooguri, {\it Comm. Math. Phys.} {\bf
   126} (1989) 49.
\bi{FGPP}
   P. Furlan, R. Paunov, A.Ch. Ganchev and V.B. Petkova,
   {\it Phys. Lett. } {\bf B267} (1991) 63.
\bi{FGPP2}
   P. Furlan, R. Paunov, A.Ch. Ganchev and V.B. Petkova,
   {\it Nucl. Phys.} {\bf B394} (1993) 665, hep-th/9201080.
\bi{Do}
    Vl.S. Dotsenko,  {\it Nucl. Phys.} {\bf B358} (1991) 547.
\bi{DF}
    V.S. Dotsenko and V.A. Fateev, {\it Nucl. Phys.} {\bf B240}
   (1984) 312; {\bf B251} (1985) 691.
\bi{FMI}
   B.L. Feigin and F.G. Malikov,  {\it Adv. Sov. Math.} {\bf
   17} (1993) 15, hep-th/9306137.
\bi{PRY}    J.L. Petersen, J. Rasmussen and M. Yu,
    {\it Nucl. Phys.} {\bf B 457} (1995) 309;
        ibid {\bf B 457} (1995) 343.
\bi{And}
    O. Andreev,  {\it Phys. Lett.} {\bf B363} (1995) 166.
\bi{AY}
    H. Awata and Y. Yamada, {\it Mod. Phys. Lett.}
        {\bf A7} (1992) 1185.
\bi{FM2}
    B.L. Feigin and F.G. Malikov,  {\it Lett. Math. Phys.}
        {\bf 31} (1994) 315;
\bi{FM}
    B.L. Feigin and F.G. Malikov,  Modular functor and
    representation theory of $\hat{sl(2)}$ at a rational level,
    q-alg/9511011.
\bi{GP}
   A.Ch. Ganchev and V.B. Petkova,
   {\it Phys. Lett. } {\bf B293} (1992) 56.
\bi{MFF}
    F.G. Malikov, B.L. Feigin and D.B. Fuks, {\it Funct. Anal.
    Prilozhen.} 20, no. 2 (1987) 25.
\bi{BPZ}
   A. Belavin, A. Polyakov and A. Zamolodchikov, {\it Nucl.
   Phys.} {\bf B241} (1984) 333.
\bi{FGPP3}
   P. Furlan, R. Paunov, A.Ch. Ganchev and V.B. Petkova,
  in: H. Osborn (Edt.), Proc. of  Workshop on Low Dimensional
  Topology and Quantum Field Theory,  Newton Institute,
  Cambridge, Sept., 1992, NATO ASI Series B, v. 315, Plenum Press,
  New York.
\bi{FGP}
   A.Ch. Ganchev and V.B. Petkova,
   {\it Phys. Lett. } {\bf B318} (1993) 77.
\bi{KR}
   A.N. Kirillov and N.Yu. Reshetikhin, Leningrad preprint, LOMI
   E-9-88 (1988).
\bi{GP89} A.Ch. Ganchev and V.B. Petkova,
   {\it Phys. Lett. } {\bf B233} (1989) 374.
\bi{Dot}
   Vl. S. Dotsenko,  Lectures on Conformal Theory, RIMS-559 (1986).
\bi{FGP1}
   P. Furlan, A.Ch. Ganchev and V.B. Petkova,
   {\it Int. J. Mod. Phys.} A6 (1991) 4859.
\bi{DF3}
    V.S. Dotsenko and V.A. Fateev, {\it Phys. Lett.} {\bf B154 }
   (1985) 291.
\bi{KPZ}
   A.M. Polyakov, {\it Mod. Phys Lett.} {\bf A2} (1987) 893;\\
   V.G. Knizhnik, A.M. Polyakov and A.B. Zamolodchikov, {\it
   Mod. Phys. Lett.}  {\bf A3} (1988) 819.
\bi{lit}
   A.P. Jucys, A.A. Bandzaitis, Theory of angular momentum in
   quantum mechanics, Mokslas, Vilnius, 1977 (in Russian).
\bi{PRYn}
   J.L. Petersen, J. Rasmussen and M. Yu,
   Fusion, crossing and monodromy in conformal field theory
   based on $sl(2)$ current algebra with fractional level,
   hep-th/9607129.


\end{thebibliography}
\end{document}